# Autumnal deep scattering layer from moored acoustic sensing in the subtropical Canary Basin

**by Hans van Haren**

Royal Netherlands Institute for Sea Research (NIOZ) and Utrecht University, P.O. Box 59, 1790 AB Den Burg, the Netherlands.
E-mail: hans.van.haren@nioz.nl

**Abstract**

An enhanced acoustic scatterer reflectance layer was observed in the bathypelagic zone around 1650 m in the subtropical NE-Atlantic Ocean for about two months during autumn. It resembles a classic pattern of diapause-resting, possibly of large zooplankton, shrimp and/or *Cyclothone*, at great depths well below any sunlight penetration, which is more commonly found at higher latitudes. The observed slow sink and rise of about 2-5 m per day into and out of this deep layer is considerably slower than the more than 1000 m per day of diel vertical migration (DVM). During the two-month period of deep scattering, DVM is observed to be greatly reduced.

**Introduction**

Vertical migration, especially diel vertical migration (DVM) is a ubiquitous phenomenon in ocean living organisms such as zooplankton, decapods, gelatinous plankton, *Euphausiacea* and particular small fish species (Hays 2003). Most commonly, organisms migrate up at dusk and return to deeper waters at dawn. DVM is thought to be triggered by a relative change in visible light intensity. However, evidence has shown that DVM also occurs in the bathypelagic deep-sea below 1000 m (Vinogradov 1961; Plueddemann and Pinkel 1989) where not a single photon of sunlight penetrates (Kampa 1970). DVM may become modulated by monthly lunar periodicities, also at great depths, and seasonal variations (van Haren 2007). The seasonal variations are thought to depend on food supply and various life stages, including a dormant resting stage 'diapause' (e.g., Baumgartner and Tarrant 2017). In general, the depth levels of diapause are deeper than DVM levels, presumably to avoid predation, commonly far below sunlight penetration and in cold ~0°C waters, as has been established mostly for high-latitude species *Calanus finmarchicus* and *C. hyperboreus* (e.g., Hirche 1991). However, the precise triggers for descent and re-emergence of diapause, including association with DVM, are still not known (Häfker 2018).

Plankton and fish research goes back centuries (Latreille 1829; Murray and Hjort 1912), with detailed species knowledge from net-tows (e.g., Badcock and Merrett, 1976; Schott and Johns 1987; Flagg and Smith 1989), which have a rather poor spatial and temporal coverage. But since the advance of acoustic scattering methods since the 1960s, originally and foremost designed for targeting fish (e.g., Dragesund and Olsen, 1965; Brekhovskikh and Lysanov 1990; Sutton et al. 2010; Sutton 2013), considerable details notably about DVM have been revealed also using acoustic Doppler current profilers (ADCPs) (Schott and Johns 1987; Flagg and Smith 1989; Plueddemann and Pinkel 1989); and have been verified with net-tows. The ocean is relatively transparent for sound and scattering reflectance of the order of 100 kHz acoustic sources is mainly a measure of large zooplankton, crustacea and fish abundance (Brekhovskikh and Lysanov 1990). Acoustics are non-invasive and may be used for continuous monitoring (Haney 1988). ADCPs in a moored fashion can directly measure the vertical speed of scatterer ensembles for prolonged periods of time (Plueddemann and Pinkel 1989). But acoustics are limited in that they cannot provide taxonomic resolution and they have difficulty in quantifying plankton biomass with a single frequency instrument (Fielding et al. 2004).

In the present study, a layer of enhanced acoustic scatterer reflectance was observed in a bathypelagic record of one and a half years of moored 75 kHz ADCP-data from the subtropical NE-Atlantic Canary Basin. In contrast with more commonly used shipborne echo-sounders, the spatial coverage of moored instrumentation is limited to a certain vertical range without horizontal coverage, and a single transmit frequency is used. Its advantage is a high temporal resolution and increased resolution in deep water. The objective was to identify acoustic echo intensity variability with time of this deep scattering layer including its relationship with the seasonal variability of DVM. Unfortunately no ground-truth data are available in the form of net-tows, for which we have to rely on previously reported observations from the area (e.g., Vinogradov 1961; Badcock and Merrett 1976; Andersen et al. 1997; Vinogradov 2005; Labatt et al. 2009).

**Materials and methods**

**ADCP mooring deployment**

Between 10 June 2006 (day 160) and 22 November 2007 (day 324+365), a 3900 m long mooring was deployed West of the island of Madeira at 33° 00.0´N, 22° 04.8´W, H = 5274 m water depth in the open Canary Basin, NE-Atlantic Ocean. The top-buoy at 1374 m held a downward looking self-contained 75 kHz Teledyne-RDI Long Ranger ADCP. It sampled 60 vertical bins of 10 m between 1400 and 2000 m depth once every 1800 s. The ADCP has four beams that are slanted at an angle of $\theta = 20°$ to the vertical.

The present vertical acoustic range is not reached by any detectable daylight (Kampa 1970) and temperatures vary between about 4 and 8°C, considerably warmer than in boreal and polar regions. The Canary Basin is the domain for the abundant presence of mesoscale eddies, notably those of Mediterranean outflow waters 'Meddies' that transport relatively warm, salty waters occasionally to 2000 m.

Typical horizontal current speeds were 0.1 m s$^{-1}$ and tidally dominated. Despite the long mooring line, pressure and tilt sensor information showed generally <1.5° tilt angle and top-buoy excursions across <1.5 m in the vertical and <100 m in horizontal directions due to current drag. These values were maximally doubled when a Meddy passed.

**Determination of vertical migration from moored ADCP-data**

The current components (u, v, w) in the associated Cartesian coordinates (x-East, y-North, z-up) are measured at different depth levels z as averages over the horizontal acoustic beam spread. This is due to the $\theta = 20°$ vertical slant angle. The spread measures 20 to 440 m, depending on the range from the ADCP. The acoustic reflections 'echo intensities' (I) are obtained per 1°-aperture beam. The echo intensities are thus averaged over horizontal (and vertical) scales of about 10 m. As the I-data are generally dominated by the attenuation of sound through the water column, the suspended particle signal from acoustic scatterer reflectance dI (in decibels dB) is obtained by subtracting the water attenuation and relative to the instrumental noise level (Gostiaux and van Haren 2010). Far from bottom boundaries, as in the open North-Atlantic Ocean, the dominant source for 75 kHz ADCP dI are species that have sizes larger than about 0.01 m. Without ocean life of these sizes (75 kHz) ADCPs do not function in clear ocean waters. Variations in dI will represent changes in species abundance, size and form (Fielding et al. 2004).

For monitoring daily variations in the relative amount of echo intensity, mean composites are computed per selected monthly period (N = 31 days),

$$dI_c(z,t) = \frac{1}{N}\sum_{n=1}^{N} dI_n(z,t), \qquad (1)$$

in which t denotes half-hour time intervals spanning one day and $dI_n$ represents dI at a given day n (Plueddemann and Pinkel 1989). Similarly, for monitoring vertical movements the composite $w_c$ is computed from w,

$$w_c(z,t) = \frac{1}{N}\sum_{n=1}^{N} w_n(z,t). \qquad (2)$$

**Results**

The seasonal cycle is obvious in the deep acoustic time series observations centered around z = -1650 m (Fig. 1a,b). At this depth, water temperatures are T ≈ 5°C, and increase to T ≈ 6°C when a Meddy passes above. The large band of enhanced relative acoustic echo intensity dI > 20 dB, i.e. a factor of 100 in magnitude, above acoustic noise level slowly descents from day 210 (end July) onward and re-ascents around day 350 (mid-December). Between about days 260 (mid-September) and 330 (end-November) the ensemble of scatterers remains around the same depth and shorter period daily variations are minimal (Fig. 1b). Before and after this two-month 'resting-period' at depth are daily variations, which are visible in the thin vertical lines and detailed below. These variations represent DVM and are particularly strong in late spring/early summer up to day 190 and, one year later, between days 510 and 570. Associated daily variations in w-amplitude were up to 0.03 m s$^{-1}$ (Fig. 1c). The speeds were maintained for several hours, resulting in a vertical range of DVM of up to 300 m (Fig. 1a), all far below the depths where sunlight penetrates. However, the speed into and from the resting depth is much smaller, below the instrumental noise level of about $|w| < 0.005$ m s$^{-1}$. Estimating from the ascent and descent in dI the speed measures about 2 to 5 m d$^{-1}$, or about 2 to 6×10$^{-5}$ m s$^{-1}$, a factor of 1000 smaller than DVM-speeds. The resting-depth scatterer abundance is not associated with the passage of a Meddy above, which is observed in horizontal currents between days 350 and 450 (Fig. 1d).

The large discrepancy between spring and autumn DVM is visible in frequency spectra that highlight sinusoidal periodic motions (Fig. 2). During late spring/early summer (days 161 to 192) DVM is entirely responsible for the diurnal '$D_1$' peak in dI-variance, which extends two orders of magnitude above noise level. The $D_1$-signal shows minor peaks at its higher harmonic frequencies $D_2$, $D_3$ etcetera, which are a result of the Fourier decomposition of a square DVM-signal into sines and cosines. The diurnal peak is not related to currents, as the kinetic energy spectrum demonstrates significant peaks at semidiurnal lunar $M_2$- and near-inertial frequency f, the latter reflecting motions due to the Earth rotation.

In contrast, the same dI-analysis during the deep resting-period in early autumn (days 265 to 296) demonstrates hardly any significant peaks above noise level, except perhaps a small $M_2$-tidal peak due to internal wave motions advecting the scatterers. During this period, the currents show more or less the same spectral information as during early summer.

The contrast between early-summer and autumn diurnal migrations is also clear in monthly mean daily composite plots in Fig. 3 using equations (1a) and (1b). In early summer in particular, the composite relative echo intensity descends between sunrise and mid-day from about -1500 m to about -1800 m and ascends well before sunset (Fig. 3a). The $dI_c$-variation is associated with vertical downward and upward velocity, respectively, with a composite mean amplitude of 0.01 m s$^{-1}$ (Fig. 3b). The timing, vertical range and vertical speed are typical for DVM well below the photic zone (e.g., Plueddemann and Pinkel 1989; van Haren 2007). Three months later in our autumn observations however, a stagnant high echo intensity layer is observed around -1650 m (Fig. 3c), with negligible vertical motion (Fig. 3d).

During the second summer/autumn period the same phenomena of strong DVM before and weak DVM during the deep resting-period are observed precisely 365 days later, although the dI now barely extends above acoustic noise level. The latter may be due to a different life stage or different abundance of the scatterers resulting in different acoustic reflectance now dominating the deep resting-period.

**Discussion**

In the deep bathypelagic waters of the subtropical Canary Basin, a large scatterer layer occurs during autumn, with a two-month strongly reduced DVM between about mid-September (days 260; 625) and late November (days 330; 685). The present observations confirm that DVM occurs in deep waters without possible sunlight cues during other periods of the year. While the variation in resting-period-, but not DVM-, acoustic scatterer reflectance is large between different years, its depth is the same around z = -1650 m in the present data. The large scatterer depth is comparable with the deepest median depth observed for subpolar North-Atlantic diapause, of *C. hyperboreus*, which can rest between z = -1000 and z = -3000 m but typically around z = -1500 m in the Arctic where temperatures are below zero (Hirche 1991, Gislason et al. 2007, Baumgartner and Tarrant 2017).

The 1000-times different vertical speeds during DVM compared with those before and after the resting-period suggest different driving mechanisms. It has been demonstrated for *C. finmarchicus*, remaining for up to 5 months in diapause in T < 0.5°C subpolar >700 m deep waters, that wax ester lipids are the drivers for the vertical diapause motions due to their thermal expansion and compressibility higher than sea water (Visser and Jónasdóttir 1999). Model results indicate ascent rates in the deep of less than 5 m $d^{-1}$, confirming the present acoustic estimates. The two month reduction in DVM during the resting-period may associate with neutral buoyancy of species during their dormant stage (Schründer et al. 2014), which requires further investigation for Canary Basin species.

The present moored observations show a layer of large scatterer reflectance resting-period to occur in subtropical regions at large depths, with moderate temperatures of about 5°C. The layer may be related to (mainly mesopelagic) *Cyclothone* that are known not to show DVM (Badcock and Merrett 1976). The large scatterer layer may also be related to organisms normally showing DVM, but being in diapause. The present autumnal period of about two months with reduced DVM is observed before the shortest day-length. The layer with reduced DVM may be related with food shortage, as the autumn period coincides with a minimum in phytoplankton chlorophyll-a found in sediment traps at 1000 and 3000 m in the Canary Basin in some years in the 1990s (Neuer et al., 1997). While DVM is reduced, it is unlikely to provide a cue to diapause, also because it is itself not directly sunlight driven. It remains to be investigated how and if DVM and resting-period endogenous clocks are related and whether the latter is driven by a circannual clock as suggested for diapausing *C. finmarchicus* (Häfker 2018). While the diel clock gene cycle in *C. finmarchicus* is found ambiguous during diapause, it may persist on an individual level. The present data suggest DVM-synchronization is resumed as soon as emergence-ascent starts after the resting-period. Such circannual periodicity may be photoperiod triggered (Goldman et al. 2004), as DVM has been associated with latitudinal and seasonal day-light variation, but at great depths without sunlight cues (van Haren and Compton 2013, Schoenle 2015).

**Compliance with ethical standards** The author declares that there are no conflicts of interest.

**Humans/animals rights** No humans and/or animals are involved in the present research.

**Acknowledgments** I thank the crew of the R/V Pelagia and NIOZ-MTM for the deployment and recovery of the moorings. The funding of instrumentation by N.W.O. large investment program Long-term Ocean Current Observations (LOCO) is gratefully acknowledged.

**References**

Andersen V, Sardou J, Gasser B (1997) Macroplankton and micronekton in the northeast tropical Atlantic: abundance, community composition and vertical distribution in relation to different trophic environments. Deep-Sea Res I 44:193-222.

Badcock J, Merrett NR (1976) Midwater fishes in the eastern North Atlantic-I. Vertical distribution and associated biology in 30°N, 23°W, with developmental notes on certain myctophids. Progr. Oceanog. 7:3-58.

Baumgartner MF, Tarrant AM (2017) The physiology and ecology of diapause in marine copepods. Ann Rev Mar Sci 9:387-411.

Brekhovskikh LM, Lysanov YuP (1990) Fundamentals of ocean acoustics, 2$^{nd}$ Edition. Springer-Verlag, Heidelberg, D, 270 pp.

Dragesund O, Olsen S (1965) On the possibility of estimating year-class strength by measuring echo-abundance of 0-group fish. FiskDir Skr Ser HavUnders 13(8): 48-75.

Fielding S, Griffiths G, Roe HSJ (2004) The biological validation of ADCP acoustic backscatter through direct comparison with net samples and model predictions based on acoustic-scattering models. ICES J Mar Res 61:184-200.

Flagg CN, Smith SL (1989) On the use of the acoustic Doppler current profiler to measure zooplankton abundance. Deep-Sea Res 36:455-474.

Gislason A, Eiane K, Reynisson P (2007) Vertical distribution and mortality of *Calanus finmarchicus* during overwintering in oceanic waters southwest of Iceland. Mar Biol 150:1252-1263.

Goldman B et al. (2004) Circannual rhythms and photoperiodism. In: Dunlap JC, Loros JJ, DeCoursey PJ (eds) Chronobiology: biological timekeeping. Sinauer, Sunderland, MA, 107-142.

Gostiaux L, van Haren H (2010) Extracting meaningful information from uncalibrated backscattered echo intensity data. J Atmos Ocean Technol 27:943-949.

Häfker NS (2018) The molecular basis of diel and seasonal rhythmicity in the copepod *Calanus finmarchicus*. PhD-thesis, Univ. Oldenburg, D, 144+XXVIII pp.

Haney JF (1988) Diel patterns of zooplankton behavior. Bull Mar Sci 43:583-603.

Hays GC (2003) A review of the adaptive significance and ecosystem consequences of zooplankton diel vertical migrations. Hydrobiol 503:163-170.

Hirche H-J (1991) Distribution of dominant calanoid copepod species in the Greenland Sea during late fall. Polar Biol 11:351-362.

Kampa EM (1970) Underwater daylight and moonlight measurements in the eastern North Atlantic. J Mar Biol Ass UK 50:397-420.

Labat J-P, et al (2009) Mesoscale distribution of zooplankton biomass in the northeast Atlantic Ocean determined with an optical plankton counter: Relationships with environmental structures. Deep-Sea Res I 56:1742-1756.

Latreille PA (1829) In: Cuvier JLNF (ed) Le règne animal distribué d'après son organisation, pour servir de base a l'histoire naturelle des animaux et d'introduction a l'anatomie compare. Déterville, Paris, F: 1-584.

Murray J, Hjort J (1912) The depths of the ocean. MacMillan, London, UK, 821 pp.

Neuer S, Ratmeyer V, Davenport R, Fischer G, Wefer G (1997) Deep water particle flux in the Canary Island region: seasonal trends in relation to long-term satellite derived pigment data and lateral sources. Deep-Sea Res I 44:1451-1466.

Plueddemann AJ, Pinkel R (1989) Characterization of the patterns of diel migration using a Doppler sonar. Deep-Sea Res 36:509-530.

Schoenle A (2015) Time to (dia)pause. Clock gene expression patterns in the calanoid copepod *Calanus finmarchicus* during early and late diapause. MSc-thesis, Univ. Bremen, D, 49 pp.

Schott F, Johns W (1987) Half-year long measurements with a buoy-mounted acoustic Doppler current profiler in the Somali current. J Geophys Res 92:5169-5176.

Schründer S, Schnack-Schiel SB, Auel H, Sartoris FJ (2014) Observations of neutral buoyancy in diapausing copepods Calanoides acutus during Antarctic winter. Polar Biol 37:1369-1371.

Sutton TT, Wiebe, PH, Madin, L, Bucklin A (2010) Diversity and community structure of pelagic fishes to 5000 m depth in the Sargasso Sea. Deep-Sea Res II 57:2220-2233.

Sutton TT (2013) Vertical ecology of the pelagic ocean: classical patterns and new perspectives. J Fish Biol 83:1508-1527.

van Haren H (2007) Monthly periodicity in acoustic reflections and vertical motions in the deep ocean. Geophys Res Lett 34:L12603. doi:10.1029/2007GL029947.

van Haren H, Compton TJ (2013) Diel vertical migration in deep sea plankton is finely tuned to latitudinal and seasonal day length. PLoS ONE 8, e64435 doi:10.1371/journal.pone.0064435.

Vinogradov ME (1961) Feeding of the deep-sea zooplankton. Rapp P V Réun Cons Int Explor Mer 153:114-120.

Vinogradov GM (2005) Vertical distribution of macroplankton at the Charlie-Gibbs Fracture Zone (North Atlantic), as observed from the manned submersible "Mir-1". Mar Biol 146:325-331.

Visser AW, Jónasdóttir SH (1999) Lipids, buoyancy and the seasonal vertical migration of *Calanus finmarchicus*. Fish Oceanogr 8:100-106.

**Fig. 1**. Overview of 1.5 years ADCP-data from the subtropical Canary Basin ranging vertically between z = -1400 and -2000 m. Time in 2007 is yearday +365. **a** Time-depth series of echo intensity relative to its background attenuation and noise level. Below z = -1850 m noise dominates due to insufficient acoustic scatterers. The horizontal bars indicate the 31 day periods of Fig. 2. **b** Time series of relative echo intensity, averaged hourly and over 100 m between [-1700, -1600] m. **c** As b., but for the vertical current component. **d** As b., but for horizontal East-West current component.

**Fig. 2**. Spectra from relative echo intensity (solid graphs) and kinetic energy (dashed graphs) averaged between z = -1600 and -1700 m for two periods of one month each. In thin graphs the summer period (thin horizontal bar in Fig. 1a) with strong diel vertical migration DVM. In bold graphs the autumn period (thick bar in Fig. 1a) of deep large scatterer reflectance with negligible DVM. Indicated on top are the solar diurnal frequency $D_1$ and its first harmonic $D_2$, inertial frequency f and semidiurnal lunar tidal frequency $M_2$.

**Fig. 3**. Monthly averaged composites of daily variations in relative echo intensity following Eq. (**1**) (panels a, c) and vertical current following Eq. (**2**) (panels b, d) for data from the two periods in Fig. 2. Data below -1890 m were too noisy to produce reliable composites (see a remnant in d. at the lower edge). Symbols 'r' indicate local sunrise and 's' local sunset. (a) Relative echo intensity for the 31 yeardays during early summer indicated above the panel, when DVM was not weak. (b) Vertical current associated with a. (c) As a., but for an autumn 31 days when DVM was negligible. (d) As b., but associated with a.

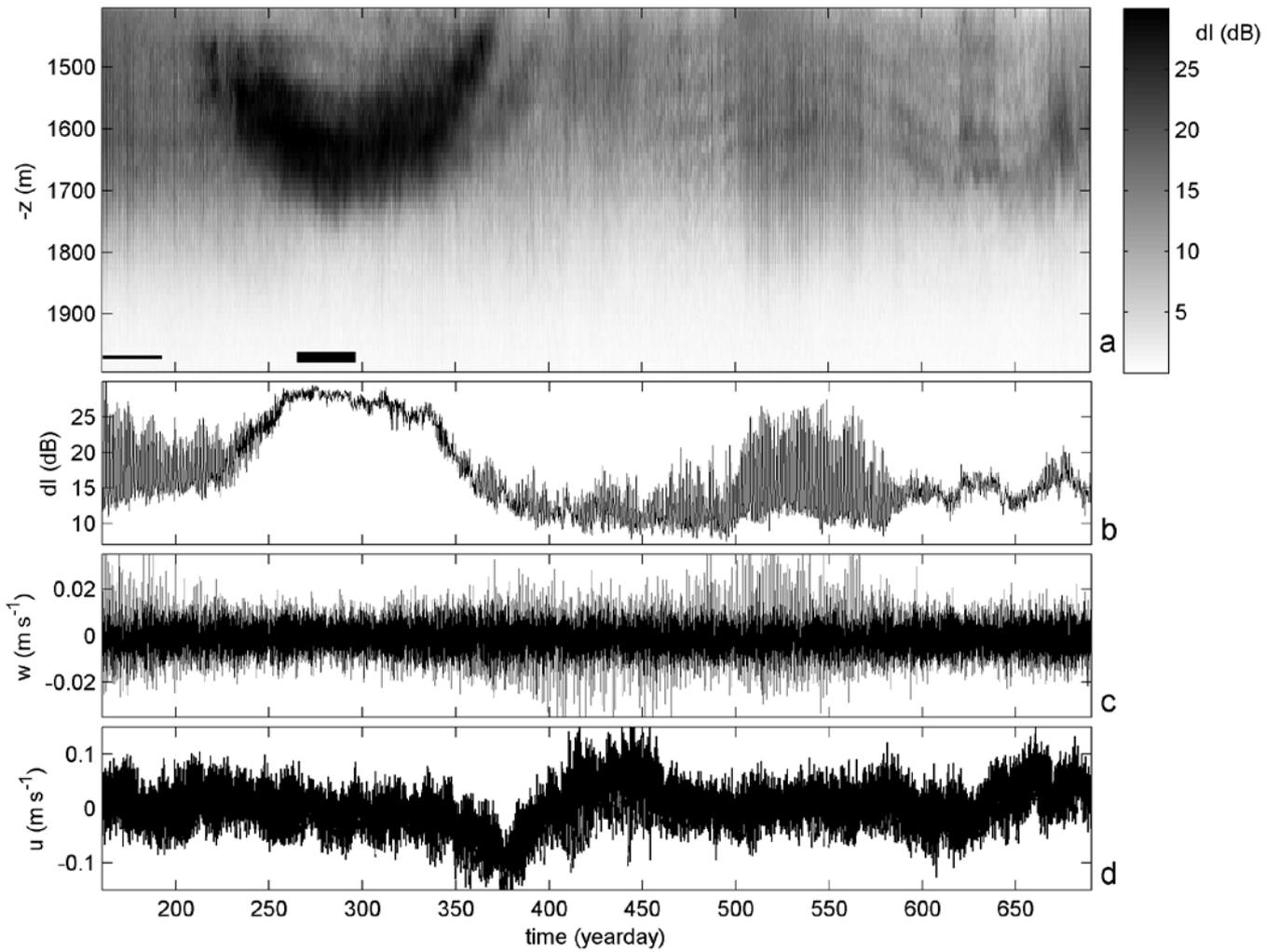

**Fig. 1**. Overview of 1.5 years ADCP-data from the subtropical Canary Basin ranging vertically between z = -1400 and -2000 m. Time in 2007 is yearday +365. **a** Time-depth series of echo intensity relative to its background attenuation and noise level. Below z = -1850 m noise dominates due to insufficient acoustic scatterers. The horizontal bars indicate the 31 day periods of Fig. 2. **b** Time series of relative echo intensity, averaged hourly and over 100 m between [-1700, -1600] m. **c** As b., but for the vertical current component. **d** As b., but for horizontal East-West current component.

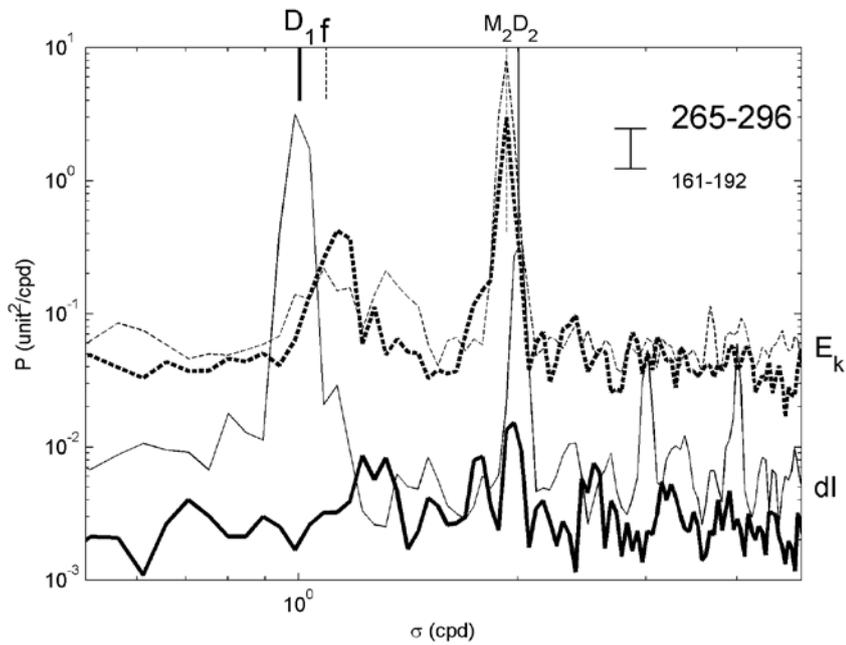

**Fig. 2**. Spectra from relative echo intensity (solid graphs) and kinetic energy (dashed graphs) averaged between z = -1600 and -1700 m for two periods of one month each. In thin graphs the summer period (thin horizontal bar in Fig. 1a) with strong diel vertical migration DVM. In bold graphs the autumn period (thick bar in Fig. 1a) of deep large scatterer reflectance with negligible DVM. Indicated on top are the solar diurnal frequency $D_1$ and its first harmonic $D_2$, inertial frequency f and semidiurnal lunar tidal frequency $M_2$.

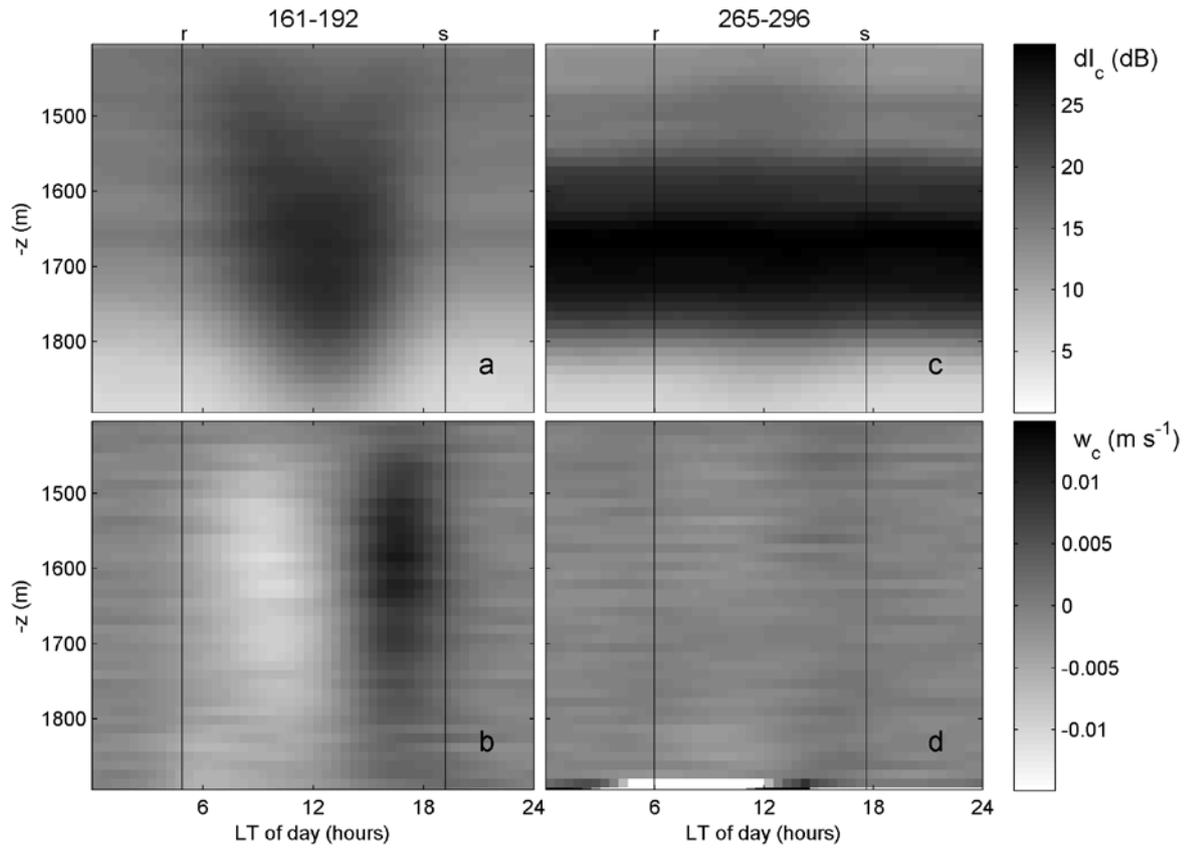

**Fig. 3**. Monthly averaged composites of daily variations in relative echo intensity following Eq. (**1**) (panels a, c) and vertical current following Eq. (**2**) (panels b, d) for data from the two periods in Fig. 2. Data below -1890 m were too noisy to produce reliable composites (see a remnant in d. at the lower edge). Symbols 'r' indicate local sunrise and 's' local sunset. (a) Relative echo intensity for the 31 yeardays during early summer indicated above the panel, when DVM was not weak. (b) Vertical current associated with a. (c) As a., but for an autumn 31 days when DVM was negligible. (d) As b., but associated with a.